\documentclass[12pt]{iopart}
\usepackage{url}
\usepackage{harvard}

\usepackage{graphicx}

\begin{document}

\title[Time-dependent determination of photodissociation and radiative association cross sections]{Determination of photodissociation and radiative association cross sections from the same time-dependent calculation}

\author{S Vranckx$^{1,2}$, J Loreau$^{1}$, M Desouter-Lecomte$^{2,3}$ and N Vaeck$^{1}$}

\address{$^1$ Service de Chimie Quantique et Photophysique, CP 160/09 Universit\'e Libre de Bruxelles, B-1050 Brussels, Belgium \\
         $^2$ Laboratoire de Chimie Physique (UMR8000), Universit\'e Paris-Sud, Orsay 91405, France \\
         $^3$ D\'{e}partement de Chimie, B6c Universit\'{e} de Li\`{e}ge, Sart Tilman, 4000 Li\`{e}ge, Belgium}

\begin{abstract}
We illustrate some of the difficulties that may be encountered when computing photodissociation and radiative association cross sections from the same time-dependent approach based on wavepacket propagation. The total and partial photodissociation cross sections from the thirty-three vibrational levels of the $b$ $^3\mathrm{\Sigma}^+$ state of HeH$^+$ towards the 9 other $^3\mathrm{\Sigma}^+$ and 6 $^3\mathrm{\Pi}$ $n$ = 2, 3 higher-lying electronic states are calculated, using the autocorrelation method introduced by \citeasnoun{Heller1978} and the method based on the asymptotic behaviour of wavepackets introduced by \citeasnoun{Balint-Kurti1990}. The corresponding radiative association cross sections are extracted from the same calculations, and the photodissociation and radiative association rate constants are determined. 
\end{abstract}

\maketitle

\section{Introduction}
\label{intro}

This work aims at illustrating various problems that may arise when one wants to compute accurate photodissociation cross sections and radiative association cross sections using the same time-dependent wavepacket propagation. The high sensitivity of the radiative association cross sections at low energy is discussed. 

While the computation of the photodissociation cross section from the lowest rovibrational level of the initial state is usually straightforward, the computation of individual cross sections for each initial rovibrational level tends not only to be time-consuming (as a full propagation is necessary for each $v'',J'' \rightarrow v',J'$ transition) but also to suffer from additional difficulties, mostly stemming from the wider spatial distribution of the initial wavepacket and from the appearance of centrifugal barriers in the excited states potentials for high values of $J'$. 
 
We illustrate some of these difficulties through a study of the photodissociation of the $b$ $^3\mathrm{\Sigma}^+$ state of HeH$^+$. Although HeH$^+$ is one of the simplest closed-shell diatomic ions in its fundamental state, its nuclear dynamics in excited states is far from trivial as they are strongly coupled by non-adiabatic interactions. In the last decades, this cation has been the subject of numerous theoretical studies motivated by its relative simplicity, which allows for high-accuracy computations \cite{Stanke2006,Miyake2011} ; its potential role as a diagnostic tool in plasma physics \cite{Rosmej2006,Loreau2010c} and its importance in astrochemistry \cite{Roberge1982,Dalgarno2005}. It is indeed predicted to be the first molecular species to have formed in the universe, by radiative association of H$^+$ and He \cite{Galli1998}, and to be abundant in certain planetary nebulae \cite{Cecchi-Pestellini1993} and Helium-rich white dwarfs \cite{Harris2004}. Every attempt of extra-terrestrial observation of this cation so far has however proven inconclusive at best. The short radiative lifetime ($\tau \approx 10^{-8}$ s \cite{Chibisov1996}) of its $b$ $^3\mathrm{\Sigma}^+$ state indicates that it is likely to quickly deexcitate towards the metastable $a$ $^3\mathrm{\Sigma}^+$ state, which could play an important role in the astrochemistry of HeH$^+$ because of its long radiative lifetime ($\tau$ = 149 s for its lowest vibrational level \cite{Loreau2010b,Loreau2013}). 

The $b$ $^3\mathrm{\Sigma}^+$  state exhibits properties which complicate the dynamics of its photodissociation. It is close in energy to a large number of other states, both bound and dissociative, which are strongly coupled by non-adiabatic interactions. Its large number of rovibrational levels also make it a good example of the importance of taking the vibrational dependence of the photodissociation cross sections into account.

\section{Molecular data}
\label{moleculardata}

\begin{figure*}
\includegraphics[width=1.00\textwidth]{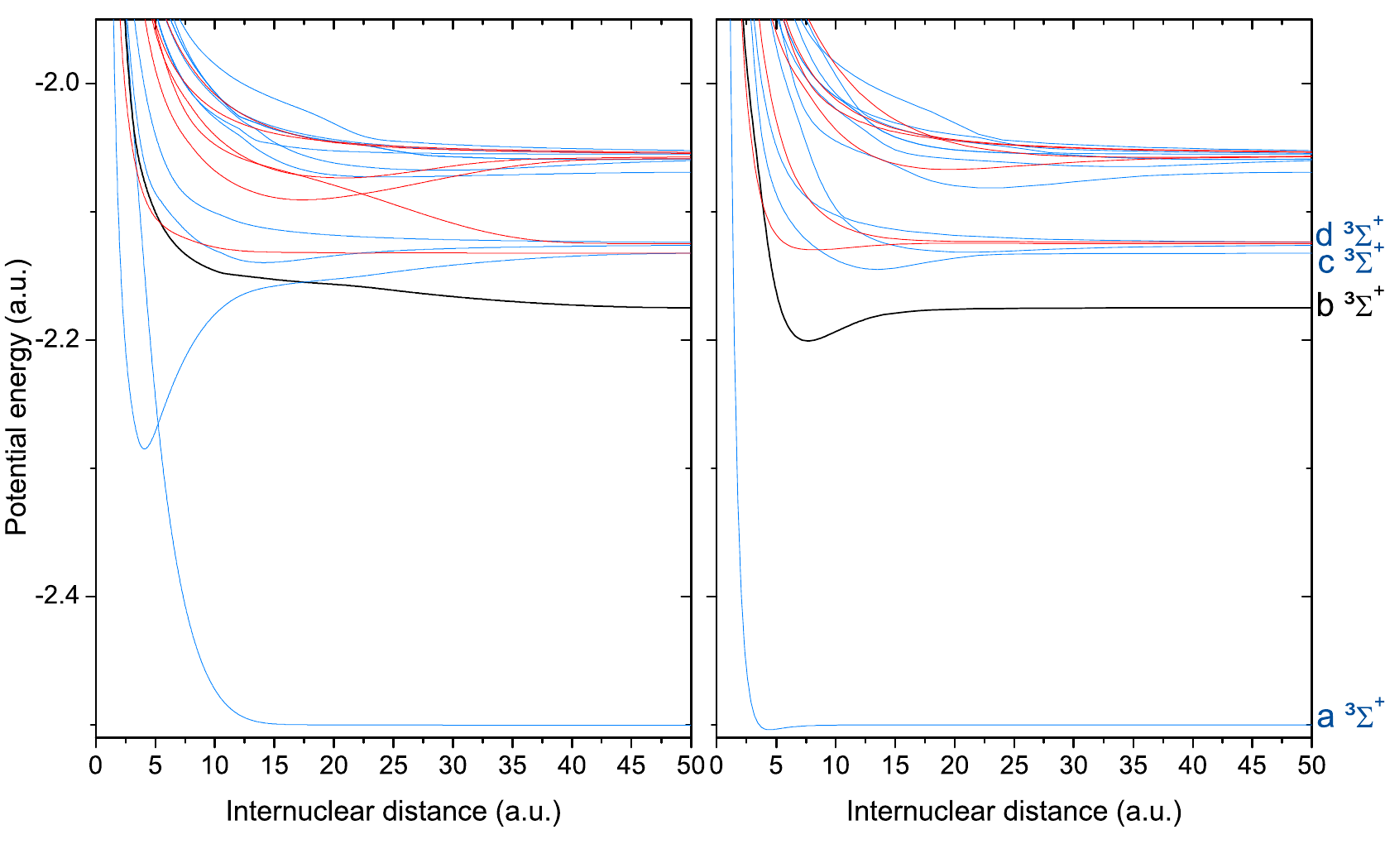}
\caption{Diabatic (left) and adiabatic (right) potential energy curves of the $n$ = 1--3 $^3\mathrm{\Sigma}^+$ (blue, black) and $^3\mathrm{\Pi}$ (red) states of HeH$^+$. The $b$ $^3\mathrm{\Sigma}^+$ state is shown in black.}
\label{pecs}      
\end{figure*}

All dynamical calculations were performed using the potential energy curves, non-adiabatic radial couplings and dipole moments computed by \citeasnoun{Loreau2010} at the CASSCF and CI level  using the MOLPRO quantum chemistry package \cite{MOLPRO}. To ensure a correct description of the excited HeH$^+$ triplet states, these calculations were performed using the aug-cc-pV5Z basis set supplemented with one contracted Gaussian function per orbital per atom up to $n$ = 4, where $n$ is the largest principal quantum number of the atomic fragments. All $\Sigma$ and $\Pi$ triplet states up to $n$ = 3 were included in our dynamical calculations, i.e. 11 $^3\mathrm{\Sigma}^+$ as well as 6 $^3\mathrm{\Pi}$ states (shown in Fig. ~\ref{pecs}), as photodissociation perpendicular to the laser polarization was shown to play an important role in the case of the fundamental state \cite{Sodoga2009}.

Among the non-adiabatic radial couplings elements $F_{mm'}= \langle {\zeta_m} | \partial_R | {\zeta_m'} \rangle$, only the couplings between successive states $F_{m,m \pm 1}$ were taken into account in the determination of the adiabatic-to-diabatic transformation matrix ${D}(R)$, like in previous studies on this system \cite{Sodoga2009,Loreau2010c}. The equation

\begin{equation}
\partial_R D(R) + F(R) \cdot D(R) = 0 
\end{equation}

\noindent was solved by continuity with the initial condition $D(\infty)=I$, in order to ensure that the adiabatic and diabatic representations are identical at large internuclear distance. The diabatic potential energy curves are the diagonal elements of the matrix $U^{\mathrm{d}} = D^{-1} \cdot U^{\mathrm{a}} \cdot D$, where $U^{\mathrm{a}}$ is the matrix of the electronic Hamiltonian $H^{\mathrm{el}}$ in the adiabatic representation. 

As the photodissociation cross sections are highly dependent on the vibrational wavefunction of the initial state, we computed the wavefunctions for the $b$ $^3\mathrm{\Sigma}^+$ state using its potential energy curve obtained at the CASSCF and CI levels using the aug-cc-PV6Z basis set (instead of aug-cc-PV5Z) supplemented with the orbitals optimized by \citeasnoun{Loreau2010}. We then resolved the vibrational nuclear equation in a basis of B-Splines to obtain the energies and wavefunctions of its vibrational levels \cite{Bachau2001,Loreau2010b,Cauet2012}. The problem was solved for all possible values of $J$ by adding the corresponding centrifugal term to the potential. The $b$ $^3\mathrm{\Sigma}^+$ state was found to support 33 bound vibrational levels and a total of 1307 rotational levels.

\section{Photodissociation cross sections}
\label{photodissociationCS}

The cross sections were extracted from time-dependent wavepacket propagations. In this approach, the initial wavepackets $\phi^{m \Lambda' J'}_{0 \Lambda'' v'' J''} (t=0)$ corresponding to the transition from the initial state $0$ towards the excited states $m$ of symmetry $\Lambda'$ are obtained by multiplying the nuclear wavefunction of the initial state $\chi_{\Lambda'' v'' J''}(R)$ by the appropriate transition dipole moments $\mu_{0 \Lambda'',m\Lambda'}(R)$: 

\begin{equation}
\phi_{0 \Lambda'' v'' J''}^{m \Lambda' J'} (R,t = 0) =  \mu_{0 \Lambda'', m\Lambda'}(R) \chi_{\Lambda'' v'' J''} (R) 
\end{equation}

The wavepackets are then propagated in time, using the split-operator method in the present case \cite{Feit1982,Alvarellos1988,Baloitcha2001}. The total photodissociation cross section $\sigma^{\mathrm{total}}_{0 v'' J'' \rightarrow \Lambda' J'}(E)$ can then be extracted from the autocorrelation function 

\begin{equation}
C_{0 v''J''}^{\Lambda' J'} (t) = \sum_m \langle \phi_{0 \Lambda'' v'' J''}^{m \Lambda' J'} (t = 0) | \phi_{0 \Lambda'' v'' J''}^{m \Lambda' J'} (t) \rangle
\end{equation}

\noindent since it is given by \cite{Heller1978}: 

\begin{equation}
\sigma^{\mathrm{total}}_{0 v'' J'' \rightarrow \Lambda' J'}(E) = 4 \mathrm{\pi} \alpha a_0^2 E \textrm{Re} \int_0^{\infty} C^{\Lambda' J'}_{0 v'' J''}(t) e^{i(E_{0 v'' J''}+ E)t/\hbar} dt,
\end{equation}

\noindent where $E$ is the photon energy, $a_0$ is the Bohr radius and $E_{0 v'' J''}$ is the energy of the considered rovibrational level of the initial state. This method only yields the total photodissociation cross section, i.e. the sum of the partial, channel-specific cross sections: 

\begin{equation}
\sigma^{\mathrm{total}}_{0 v''J'' \rightarrow \Lambda' J'}(E) = \sum_m \mathrm{\sigma}^{\mathrm{partial}}_{0 v''J'' \rightarrow m \Lambda' J'}(E)
\end{equation}

The total cross section characterizes the depopulation of the initial state caused by the absorption of a photon rather than the dissociation of the molecule into specific fragments. It is therefore unsuitable for the computation of branching ratios or of radiative association cross sections from a specific excited channel towards the initial state (see Section \ref{radiativestabilizationCS}). 

The cross sections exhibit predissociation resonances caused by the non-adiabatic crossing between a bound state and a lower dissociative channel: trapped wavepackets move back and forth in the potential energy well of the bound state until they non-adiabatically cross to the dissociative channel, leading to the appearance of resonances in its spectrum \cite{Balakrishnan1999}.  

We chose to neglect them completely in the present calculations for several reasons. Their exact determination would have required impractically long propagation times, as a non-negligible fraction of the initial wavepacket was still found in the potential energy wells after propagation times as high as $50 \times 10^6$ a.u. of time ($1.2$ ns). Moreover, taking these resonances into account may lack physical sense if one wants to extract photodissociation and radiative association cross sections from the same wavepackets propagations, as discussed in Section \ref{radiativestabilizationCS}. Furthermore, including very long-lived resonances in our photodissociation cross sections may be unrealistic as electronic transitions caused by spontaneous radiative deexcitation, collisional processes or photoexcitation are likely to take place on shorter timescales. Finally, the impact of Fano-type resonances on the photodissociation rate constants is probably small (see Section \ref{rateconstants}). Such resonances may be more adequately described using partitioning techniques \cite{Desouter-Lecomte1997} or the time-independent formalism \cite{Dishoeck2011,Desouter-Lecomte1995}.

\begin{figure*}
\includegraphics[width=1.00\textwidth]{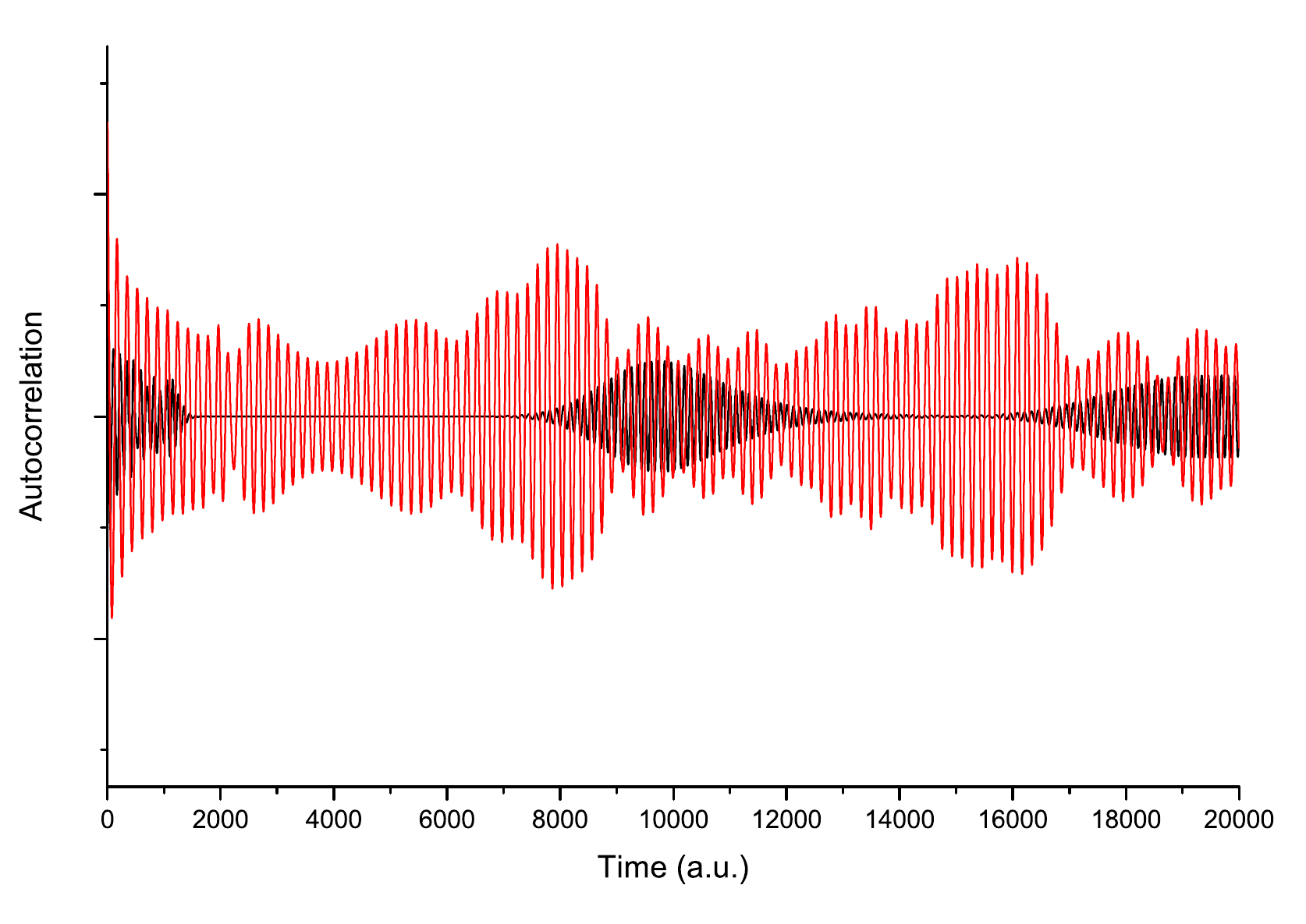}
\caption{Autocorrelation functions for the photodissociation of the $v$ = 5 (black) and $v$ = 15 (red) levels of the $b$ $^3\mathrm{\Sigma}^+$ state towards higher excited $^3\mathrm{\Sigma}^+$ states. In the case of $v$ = 5, two different time scales can clearly be observed (with the direct photodissociation contribution appearing before 2000 a.u. of time), unlike in the case of $v=15$.} 
\label{autocorrelation}      
\end{figure*}

Predissociation resonances can easily be suppressed when computing the total photodissociation cross section through the autocorrelation method for low values of $v$: their appearance occurs on a longer timescale than direct photodissociation, as illustrated by the autocorrelation function in the case of the photodissociation of the $v$ = 5 level (Fig. \ref{autocorrelation}, black curve). The departure of the wavepackets from the Franck-Condon region and the return of their trapped fractions occur on two distinct timescales and, if the propagation is stopped before their return, the computed cross section shows no sign of resonances \cite{Heller1978} and exhibits instead a smooth continuation of the cross section below the threshold energy. This can be understood intuitively as the autocorrelation function then contains no information about the trapped fraction of the wavepackets and behaves as if the totality of the wavepackets could directly reach the asymptotic region \cite{Balakrishnan1999}. This smooth envelope gradually disappears for longer propagation times as the Feshbach resonances appear in its stead. While a resonance-free total cross section can easily be obtained for the first few vibrational levels by using short propagation times, it ceases to be applicable for higher vibrational levels as the Franck-Condon region becomes wider, keeping the autocorrelation function from ever falling to zero (Fig. \ref{autocorrelation}, red curve). 

Partial photodissociation cross sections can be computed using the method introduced by \citeasnoun{Balint-Kurti1990}. In that approach, the cross sections are given by the Fourier transform of the component of the wavepackets reaching an internuclear distance $R_\infty$ located in the asymptotic region:

\begin{equation}
\mathrm{\sigma}^{\mathrm{partial}}_{0 v''J'' \rightarrow m \Lambda' J'}(E) = \frac{4 \mathrm{\pi}^2 \alpha a_0^2 k_m}{\mu} E | A^{0 v''J''}_{\Lambda' J'}(E) |^2
\end{equation}
 
\noindent with

\begin{equation}
A^{0 v''J''}_{m \Lambda' J'} (E) = \frac{1}{\sqrt{2\mathrm{\pi}}} \int_0^{\infty} \phi_{m \Lambda' J'}^{0 v'' J''} (R_\infty,t) e^{i(E_{0 v''J''}+E) t / \hbar} dt
\end{equation}

\noindent where $\mu$ is the reduced mass of the molecule, $\alpha$ is the fine structure constant and $k_m = \sqrt{2\mu(E_{0 v''J''} + E - E^m_{\mathrm{asymptotic}})}$  is the magnitude of the wave number in the considered channel $m$. 

The propagations were performed on a spatial grid of $2^{13}$ points that spans from an internuclear distance $R_\mathrm{min} = 0.1$ a.u. up to $R_\mathrm{max} = 200$ a.u. The use of a long spatial grid was made necessary by the components of the wavefunctions at large internuclear distance for the highest initial vibrational levels. The Fourier transform was performed at $R_\infty = 175$ a.u. and we placed a quadratic optical potential starting at $R = 180$ a.u. in order to avoid reflections of the wavepackets at the edge of the grid. Since this method requires the wavepackets to reach the asymptotic region, it necessitates longer propagation times than the computation of the total photodissociation cross section through their autocorrelation function. In the present work, propagations up to $5 \times 10^7$ atomic units of time ($1.2$ ns) were performed with a time step of 1 a.u. for all 33 initial vibrational levels. Tests with shorter time steps were carried out to ensure accuracy.

\begin{figure*}
\includegraphics[width=1.00\textwidth]{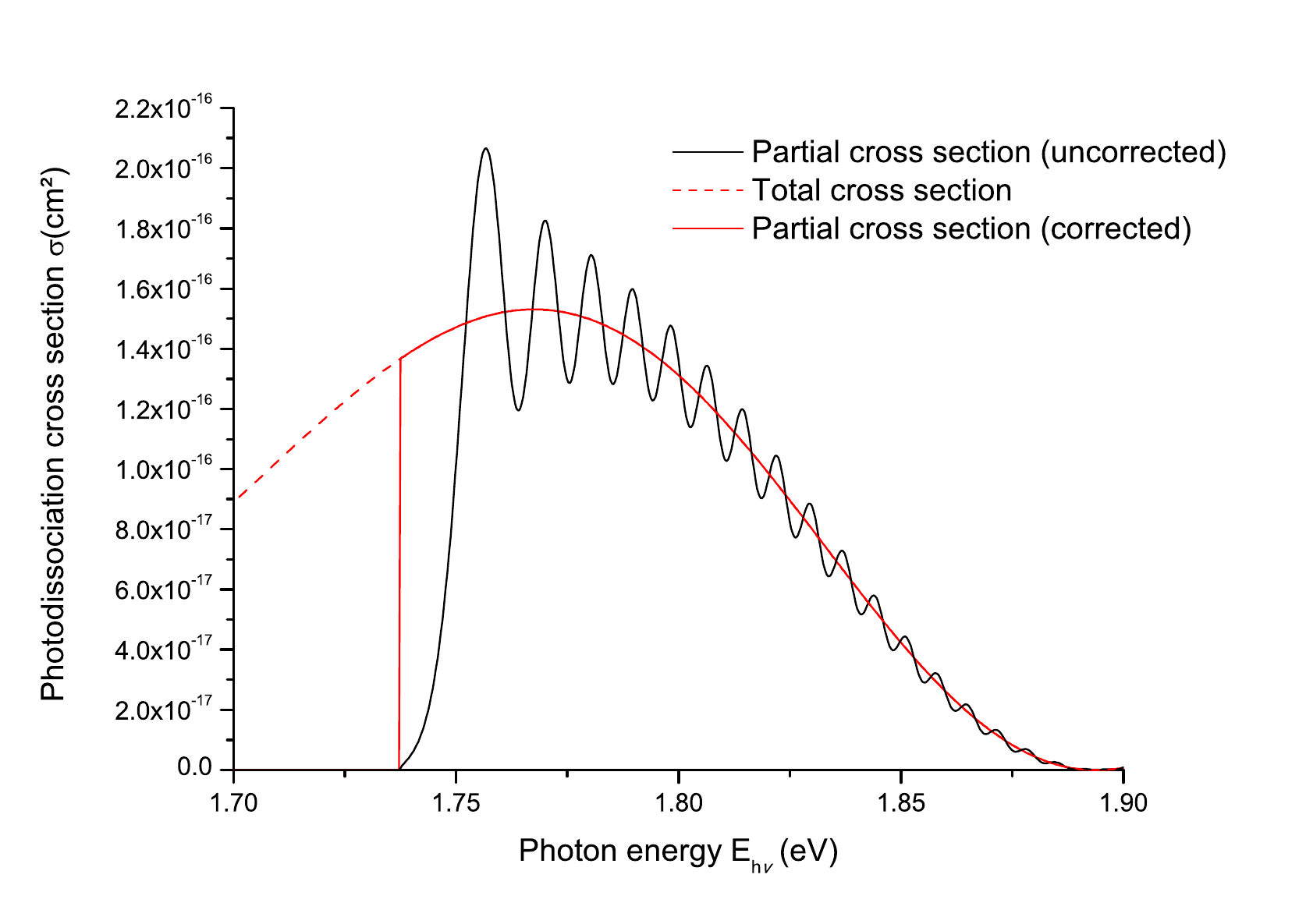}
\caption{The Gibbs phenomenon leads to the appearance of discontinuities near the threshold energy in the partial cross section from the $v$ = 2 level of the $b$ $^3\mathrm{\Sigma}^+$ state towards the H$^+$ + He(1$s$2$p$ $^3P^\mathrm{o}$) fragments (full line, black), while the total cross section as obtained through the Fourier transform of the autocorrelation function for short propagation times is perfectly smooth (dashed line). Thanks to this, the correct form of the partial cross section can be deduced (full line, red). } 
\label{gibbs}      
\end{figure*}

Unlike the autocorrelation-based approach, this method yields state-specific information. Moreover, it only takes into account the parts of the wavepackets that reach the asymptotic region of the dissociation channels and ignores their trapped fractions as long as they stay in the potential energy wells. As a consequence of this, the partial cross sections obtained with this method often start abruptly at the threshold energy and are thus affected by the Gibbs phenomenon \cite{Jerri1998}, which causes ringing artifacts around jump discontinuities in Fourier transforms, as illustrated in Fig. \ref{gibbs}.

\begin{figure*}
\includegraphics[width=1.00\textwidth]{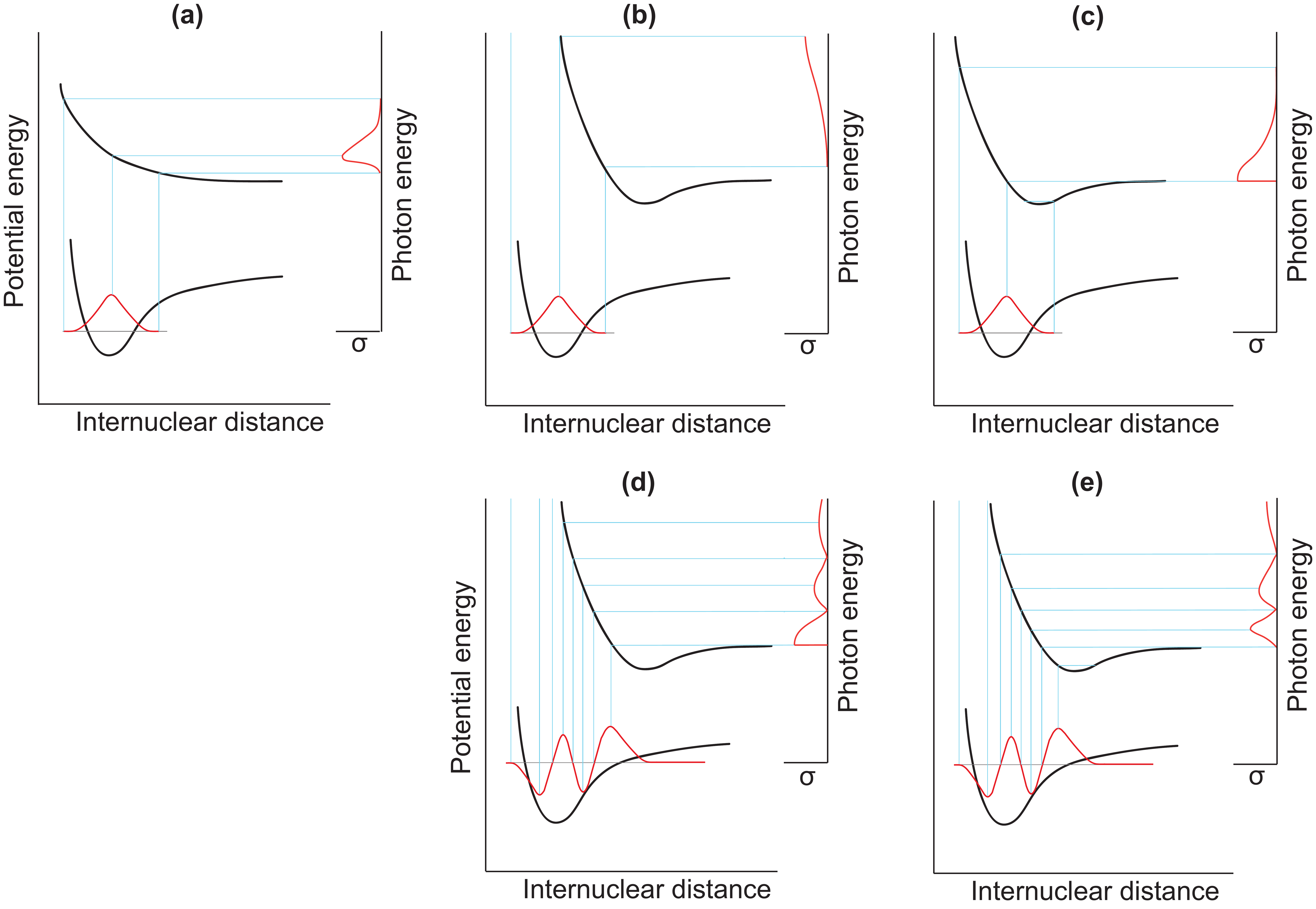}
\caption{Schematic illustration of various cases where the partial photodissociation cross section $\sigma$ does not start abruptly at the threshold energy (a, b, e) and where it does (c, d), depending on the shapes and relative position of the potential energy curves and on the initial vibrational level.} 
\label{gibbs_reflex}      
\end{figure*}

Such Gibbs oscillations only appear if the cross section is non-negligible at the threshold energy. The reflection principle allows for an intuitive understanding of the conditions under which this will happen: the dissociation channel has to be a bound electronic state, and its bound part needs to be in a region where the wavefunction of the initial rovibrational state is non-negligible (Fig. \ref{gibbs_reflex}, cases (c) and (d)). This is more likely to be the case for excited vibrational levels because of the wider spread of the vibrational wavefunction (Fig. \ref{gibbs_reflex}, compare cases (b) and (d)). Despite this, the cross section may not start abruptly at the threshold energy for certain specific values of $v''$ as one of the nodes of the wavefunction may coincide, through the reflection principle, with the asymptotic energy of the dissociation channel (Fig. \ref{gibbs_reflex}, case (e)). Whether or not Gibbs oscillations appear thus depends on several factors: the topology of the potential energy curves, their relative positions and the initial vibrational level. 

\begin{figure*}
\includegraphics[width=1.00\textwidth]{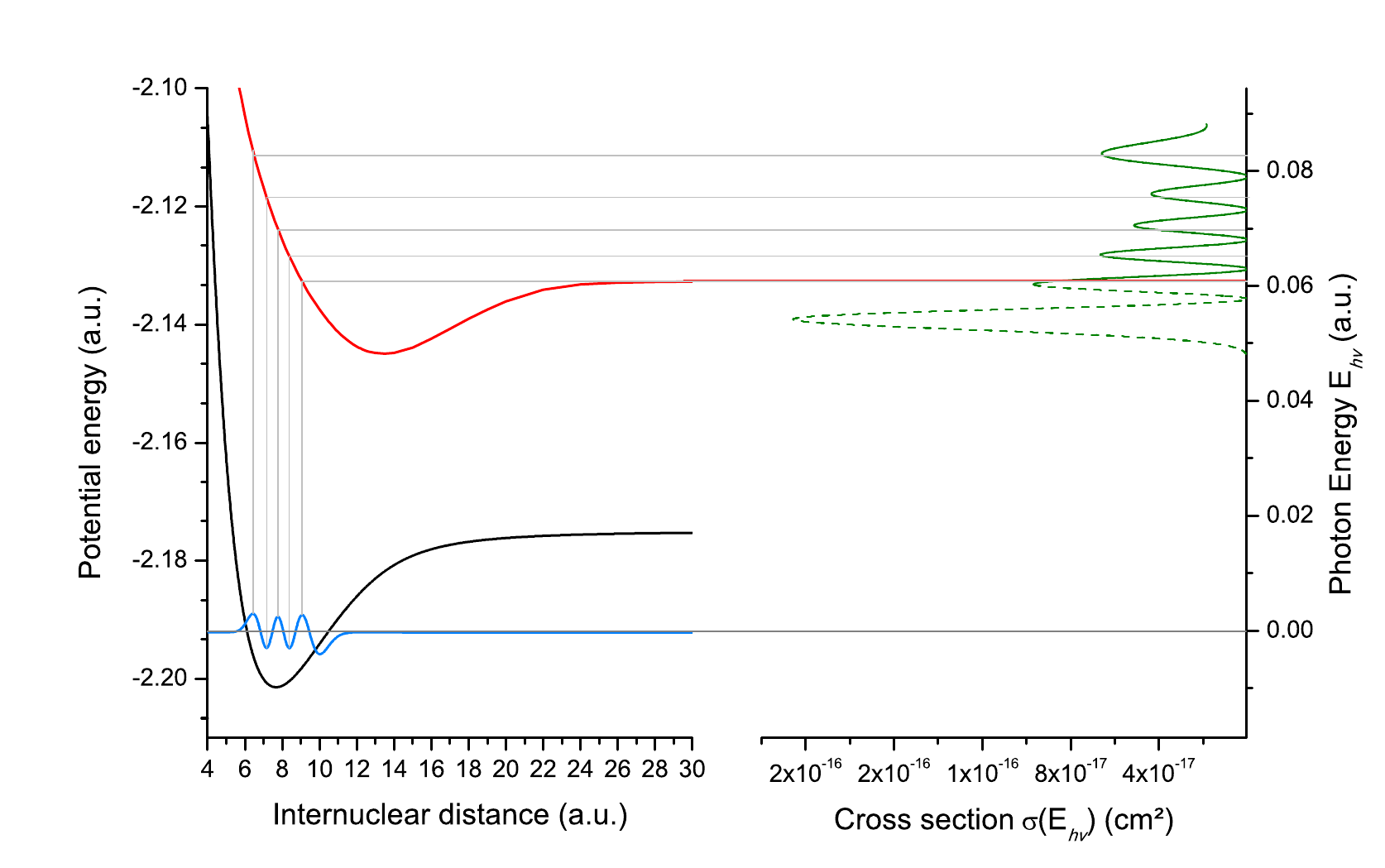}
\caption{Illustration of the reflection principle in the photodissociation cross section of the $v$ = 5 level of the $b$ $^3\mathrm{\Sigma}^+$ state (black) towards the H$^+$ + He(1$s$2$p$ $^3P^o$) channel (red). The Fourier transform of the autocorrelation not only yields the direct photodissociation cross section (green, full line), but also a contribution from predissociation which incorrectly appears as a smooth envelope for short propagation times (green, dotted line).
}
\label{reflexion}      
\end{figure*}

Although various filtering methods may be used to reduce these ringing artifacts \cite{Jerri1998}, it is generally impossible to suppress them completely. It is however possible to do so in the present case using the total photodissociation cross section calculated from the autocorrelation function for short propagation times (Fig. \ref{reflexion}): since it (incorrectly) describes predissociation as a smooth envelope, no jump discontinuity occurs at the photodissociation treshold energy. It is therefore unaffected by the Gibbs phenomenon (Fig. \ref{gibbs}). As the sum of the partial cross sections must be equal to the total cross section in the energy ranges where it contains no contribution from the incorrectly-described predissociation, the total cross section obtained for short propagation times can thus be used as a reference to suppress the ringing artifacts in a given partial cross section for a channel $m'_f$: 
 
\begin{equation}
\mathrm{\sigma}^{\mathrm{partial}}_{0 v''J'' \rightarrow m_f \Lambda' J'}(E) = \mathrm{\sigma}^{\mathrm{total}}_{0 v''J'' \rightarrow \Lambda' J'}(E) - \sum_{m' \not= m'_f} \mathrm{\sigma}^{\mathrm{partial}}_{0 v''J'' \rightarrow m' \Lambda' J'}(E)
\end{equation}

However, this is only possible if the ringing artifacts do not occur in the same range of energy for two different exit channels, as their respective contributions to the total cross sections cannot be separated in that case. By combining the results of our computations of the partial and total photodissociation cross sections, we were able to reduce or suppress the ringing artifacts near the threshold energy and to obtain the partial photodissociation cross sections from the 33 vibrational states of the $b$ $^3 \mathrm{\Sigma}^+$ state through the 15 other $n$ = 2--3 $^3 \mathrm{\Sigma}^+$ and $^3 \mathrm{\Pi}$ dissociation channels.  

The impact of the vibrational excitation of the initial state on the photodissociation cross sections is twofold: the nodal structure of the wavefunction directly affects the photodissociation cross sections (compare Fig. \ref{gibbs_reflex} (b) and (d)), moreover, the spread of the wavefunction towards high values of $R$ tends to favor transitions towards different excited channels. 

Partial photodissociation cross sections free from Feshbach resonances can often be obtained since the fraction of the wavepackets that dissociates directly usually reaches the asymptotic region on a much shorter timescale than the fraction that stays momentarily trapped in potential energy wells.  This is however not always true when considering the photodissociation of excited vibrational levels. 

Shape resonances caused by tunneling through the centrifugal barriers are also expected to appear for high values of $J'$, but the very large number of rotational levels of the $b$ $^3 \mathrm{\Sigma}^+$ state precluded us from computing individual cross sections for each of them  as a full wavepacket propagation would be necessary for each of the $J'' \rightarrow J'=J''\pm 1$ transitions towards other $^3\mathrm{\Sigma}^+$ states and $J'' \rightarrow J'=J'',J''\pm 1$ transitions towards $^3\mathrm{\Pi}$ states.  We therefore chose to neglect the rotational dependence of the cross sections, as its impact on the cross sections was shown to be smaller than that of the vibrational dependence in the case of the $X$ $^1 \mathrm{\Sigma}^+$ state \cite{Sodoga2009,Miyake2011}. However, although the envelopes of the cross sections may not vary significantly with $J''$ and $J'$, shape resonances have been shown to significantly augment photodissociation and radiative association rate constants in some systems, especially at low temperature \cite{Antipov2009}. It is common to compute the envelope of the cross section through time-dependent methods without regards to shape resonances as their exact determination would require an impractically fine energy grid and long propagation times. They can indeed be computed separately, e.g. through the Breit-Wigner formalism \cite{Breit1936,Bain1972}.

\section{Radiative association Cross Sections}
\label{radiativestabilizationCS}

Radiative association cross sections are most commonly obtained through time-independent calculations, which require the discretization of the energy continuum and separate calculations for each channel. By contrast, Martinazzo and Tantardini proposed a wavepacket-based method of computation of the radiative association cross sections, motivated by the superior scaling properties of time-dependent methods with regards to the size of the system, the number of channels considered and the energy range studied: a single time-dependent calculation yields the radiative association cross section towards all channel and for several energies at once \cite{Martinazzo2005}.

We chose a different time-dependent approach, which retains these advantages: since radiative association is the inverse process of photodissociation, the transition dipole matrix elements $M^2_{0 v'' J'', m v' J'} = | \langle \mathrm{\Psi}_{0 v'' J''} | \mu | \langle \mathrm{\Psi}_{m v' J'} \rangle |^2$ involved in both processes are the same, with the photodissociation cross section from the initial level $v'', J''$of the $0$ state being given by \cite{Barinovs1999}

\begin{equation} \label{eqphotodisso}
\mathrm{\sigma}^{\mathrm{Photodiss., partial}}_{0 v''J'' \rightarrow m}(E_{h\nu}) = \frac{8}{3} \frac{\pi^3}{c \hbar \, (4 \pi \epsilon_0)} \sum_{J'=J''-1}^{J''+1} \nu S_{J'',J'} M^2_{0 v'' J'', m v' J'}(E) 
\end{equation}
  
\noindent and the corresponding radiative association cross section from the channel $m$ towards all $v'', J''$ levels of the $0$ state being given by \cite{Barinovs1999}

\begin{equation}\label{eqradstab}
\mathrm{\sigma}^{\mathrm{Rad. Stab.}}_{m J' \rightarrow 0}(E_k) = \frac{64}{3} \frac{\pi^5}{g_m \, c^3 \hbar \, (4 \pi \epsilon_0)} \sum_{v'', J''} \sum_{J'=J''-1}^{J''+1} \frac{\nu^3_{v'' J''}}{k^2_j} S_{J'', J'} M^2_{0 v'' J'', m v' J'}(E) 
\end{equation}

\noindent where $S_{J'',J'}$ are the H\"oln-London factors and $g_m$ is the degeneracy factor of the considered channel. 

The computation of  both partial photodissociation and radiative association cross sections thus comes down to the determination of the same transition dipole matrix element $M^2_{0 v'' J'', m v' J'} = | \langle \mathrm{\Psi}_{0 v'' J''} | \mu | \langle \mathrm{\Psi}_{m v' J'} \rangle |^2$  between a bound and a free state. The radiative association cross section can thus directly be obtained from the same wavepacket propagation as the corresponding photodissociation cross section or from the photodissociation cross section itself. By comparing Eq. (\ref{eqphotodisso}) and (\ref{eqradstab}), it is indeed seen that the radiative association cross section $\mathrm{\sigma}^{\mathrm{Rad. Stab.}}_{J' \rightarrow v'',J''}$ from an excited channel $m$ towards the $v'', J''$ level of a lower state is related to the cross section characterizing the photodissociation from that $v'', J''$ level of the lower state\footnote{For clarity, we use the following convention: $v''$, $J''$ denote the rovibrational level of the lower electronic state (i.e. the initial state in photodissociation but the final state in radiative association) while  $v'$, $J'$ denote the rovibrational level of the upper state.} towards the channel $m$ by the relation \cite{Puy2007}:

\begin{equation}\label{phototostab}
\mathrm{\sigma}^{\mathrm{Rad. Stab.}}_{m v' J' \rightarrow 0 v'' J''} = \frac{E^2_{h\nu}}{\mu c^2 E_k} \mathrm{\sigma}^{\mathrm{Photodiss., partial}}_{0  v'' J'' \rightarrow m v' J'}
\end{equation}
 
While the photodissociation cross sections are expressed as a function of the incident photon energy $E_{h\nu}$, the radiative association cross sections are a function of the relative kinetic energy $E_{k}$ of the two colliding fragments. The two energy scales differ simply by the photodissociation threshold energy $E_{\mathrm{thresh}, v'' J'' \rightarrow m}$, i.e. the energy difference between the initial level $v'' J''$ and the asymptotic energy of a given fragmentation channel $m$:

\begin{equation} \label{ethreshold}
E_{k, v'' J'' \rightarrow m} = E_{h\nu} - E_{\mathrm{thresh}, v'' J'' \rightarrow m}
\end{equation}
 
The presence of the collision energy $E_k$ in the denominator of Eq. \ref{phototostab} implies that radiative association cross sections at low energy, and therefore the rate constants at low temperature, are highly sensitive to the value of the corresponding photodissociation cross sections near the threshold energy. The suppression of ringing artifacts is thus important to the determination of accurate radiative association cross sections and in cases where the corresponding photodissociation cross sections are non-negligible near the treshold energy, as illustrated in Fig. \ref{gibbs_reflex}, the radiative association cross sections will be larger for low collision energies. 

Although computing a radiative association cross section through the propagation of \emph{dissociative} wavepackets may seem counterintuitive, it presents some advantages: it circumvents the problem of the choice of the shape of the initial wavepacket in collisional problems and it allows the determination of the cross section on its whole range of energy through a single propagation, whereas several propagations are necessary to cover a wide range of energy in a time-dependent collisional approach \cite{Martinazzo2005}. Note, however, that the dissociation approach used here describes the radiative association from several collision channels towards a single bound state, whereas the usual collisional approach describes the radiative association from a single collision channel towards several lower states at once. Our approach is more time-efficient if one wants to study the formation of a molecule in a specific state by radiative association from several collision channels at once. The usual collisional approach is however preferable if one wants to study the different bound electronic states that can be formed by radiative association from a specific initial channel. 

Radiative association can occur towards any of the rovibrational level of the inferior state, its cross section $\mathrm{\sigma}^{\mathrm{Rad. Stab.}}_{m \rightarrow 0}$ is thus obtained by summing the contributions of all the rovibrational levels $\mathrm{\sigma}^{\mathrm{Rad. Stab.}}_{m v' J' \rightarrow 0 v'' J''}$. Since the rotational dependence of the cross sections was neglected in the present study, we approximated the radiative association cross sections by summing the cross sections for all vibrational levels multiplied by the corresponding number of rotational levels. 

\begin{figure*}
\includegraphics[width=1.00\textwidth]{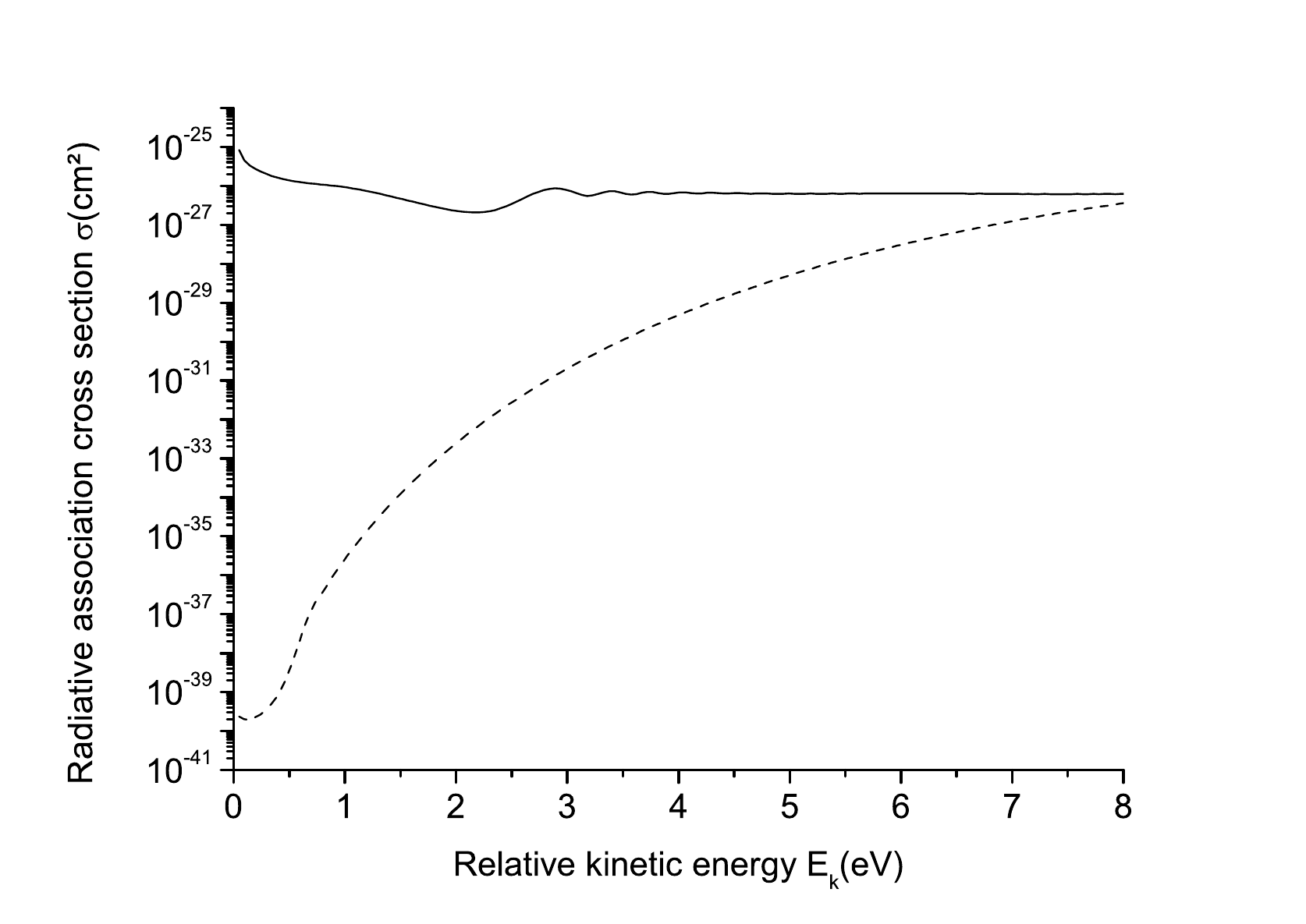}
\caption{Cross section for the radiative association of He$^+$ (1$s$) and H(1$s$) towards the fundamental  $X$ $^1 \mathrm{\Sigma}^+$ state via the $A$ $^1 \mathrm{\Sigma}^+$ channel, with (full line) and without (dotted lines) taking its vibrational dependence into account.}
\label{radx}      
\end{figure*}

We tested our method on the well-studied case of the photodissociation and radiative association between the $X$ $^1 \mathrm{\Sigma}^+$ and $A$ $^1 \mathrm{\Sigma}^+$ states of HeH$^+$ \cite{Zygelman1990,Kraemer1995,Miyake2011}. In this specific case, we found the neglect of the vibrational dependence in photodissociation to be a very bad approximation if one uses them to compute the corresponding radiative association cross sections (Fig. \ref{radx}): since the photodissociation cross section for $v''$ = 0 is very small near the threshold energy for the $A$ $^1 \mathrm{\Sigma}^+$ state ($\approx$ 13 eV, while the maximum of the cross section is located around 25 eV) \cite{Roberge1982}, its contribution to the radiative association cross section at low energy is particularly small. The largest contributions to the radiative association cross section actually come from the transitions involving the $v''$ = 7 and $v''$ = 8 levels of the $X$ $^1 \mathrm{\Sigma}^+$ state, as the corresponding photodissociation cross sections are large near the threshold energy.

\begin{figure*}
\includegraphics[width=1.00\textwidth]{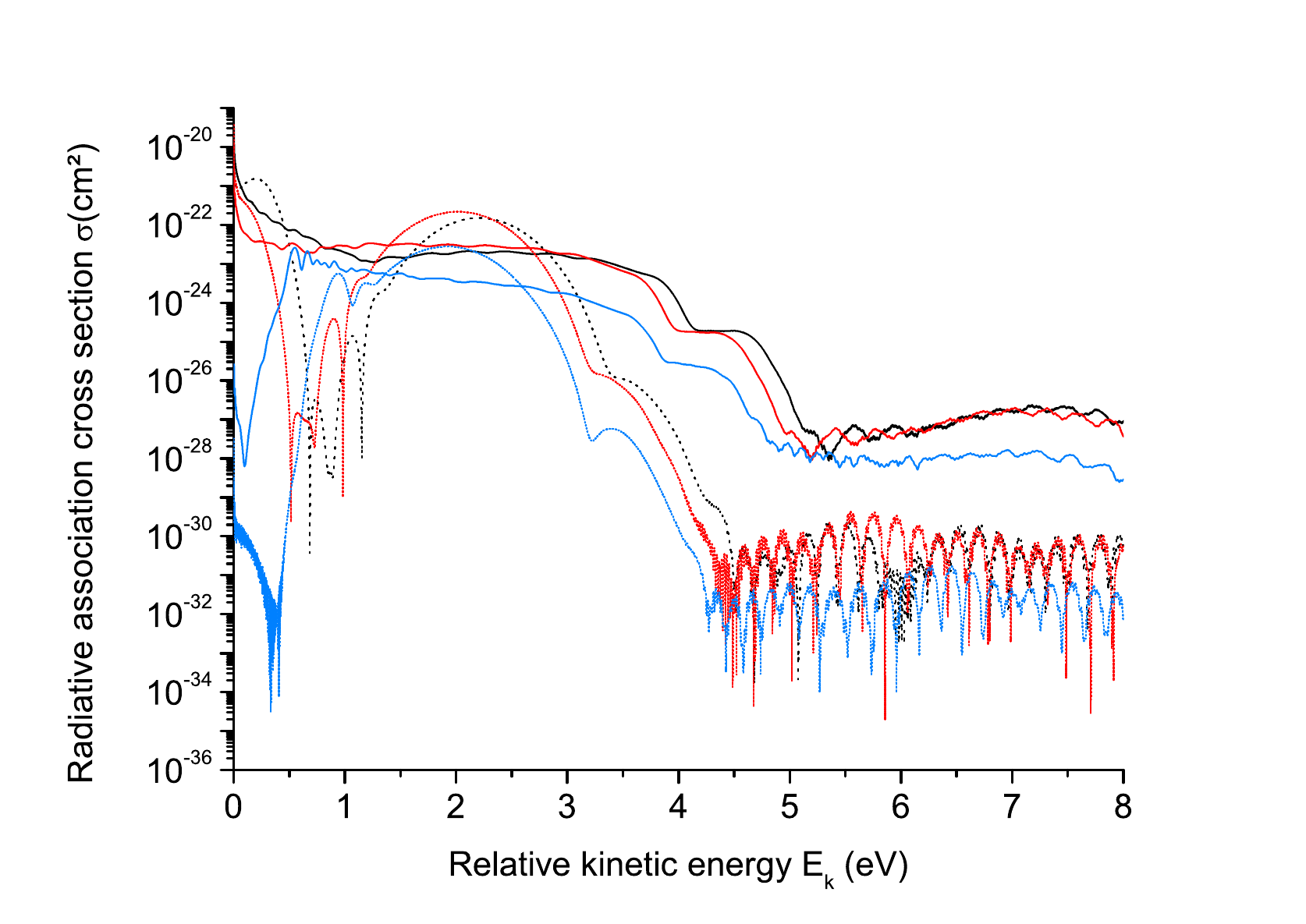}
\caption{Cross sections for the radiative association of He($1s3s$ $^3S$) + H$^+$ (black curves), He($1s3p$ $^3P^o$) + H$^+$ (red curves), and He($1s3d$ $^3S$) + H$^+$ (blue curves) towards the $b$ $^3\Sigma^+$ state of HeH$^+$ with (full lines) and without (dotted lines) taking the vibrational dependence of the cross sections into account.}
\label{rates_radstab}      
\end{figure*}

While not as dramatic, the vibrational dependence also has a significant impact in the case of the radiative association of the $b$ $^3 \mathrm{\Sigma}^+$ state (Fig. \ref{rates_radstab}), modifying the cross sections by several orders of magnitude. 

Our radiative association cross sections were obtained by suppressing the Feshbach resonances, as the non-adiabatic transitions they are caused by are unlikely to occur in the course of a collision. This can be understood intuitively in the time-dependent approach we use: during photodissociation, wavepackets trapped in a bound excited state keep oscillating in its potential energy well until they non-adiabatically cross to the continuum of a lower-lying state (provided no other phenomenon takes place, such as collisions or radiative decay). The non-radiative transition thus has repeated chances to occur, whereas it only has a small probability of taking place during a collisional process such as radiative association.

\section{Rate constants}
\label{rateconstants}

The rate constants for the photodissociation of the initial state $0$ caused by photons emitted by a blackbody of temperature $T_\star$ and of radius $R_\star$, situated at a distance $R$, is given by \cite{Roberge1982}: 

\begin{equation}
k^{\mathrm{P}}_{0 \rightarrow \Lambda'} (T) = \frac{4\pi}{h^3 c^2} \left( \frac{R_*}{R} \right)^2 \int \frac{\mathrm{\sigma}^{\mathrm{P,tot}}_{0 \rightarrow \Lambda'}(E_{h\nu}, T) E_{h\nu}^2}{e^{E_{h\nu}/k_B T_*}-1} dE_{h\nu}
\end{equation}

\noindent where $\mathrm{\sigma}^{\mathrm{P,tot}}_{0 \rightarrow \Lambda'}(E_{h\nu}, T)$ is the total photodissociation cross section characterizing the destruction  of the $0$ state assuming a Maxwell-Boltzmann distribution of population in the rovibrational levels: 

\begin{equation}
\mathrm{\sigma}^{\mathrm{P, tot}}_{0 \rightarrow \Lambda'}(E_{h\nu}, T) =  \frac{\sum_{v'', J''} (2 J'' + 1) \; e^{- E_{v'',J''}/k_B T} \mathrm{\sigma}^{\mathrm{total}}_{0 v''J'' \rightarrow \Lambda' J'}(E_{h\nu})} {\sum_{v'', J''} (2 J'' + 1) \; e^{- E_{v'',J''}/k_B T}}
\end{equation}

Note that $T$ stands for the temperature of the Maxwell-Boltzmann distribution of population in the initial state while $T_*$ denotes the temperature of the blackbody emitting the photons responsible for the photodissociation process.

As in Roberge and Dalgarno's study of the photodissociation of the $X ^1\mathrm{\Sigma}^+$ state of HeH$^+$, we chose a ratio $(R_\star/R)^2$ of 10$^{-13}$, which corresponds to the physical conditions met in planetary nebulae such as NGC7027. As previously mentioned, the cross sections were assumed to be independent of $J''$ (i.e. $\sigma_{v,J''}=\sigma_{v,J''=0}$ for all $J''$). The impact of the Fano-type resonances on the photodissociation rate constants was not assessed, but it is expected to be small, as their asymmetric profile would likely limit their net result upon integration of the cross sections.

\begin{figure*}
\includegraphics[width=1.00\textwidth]{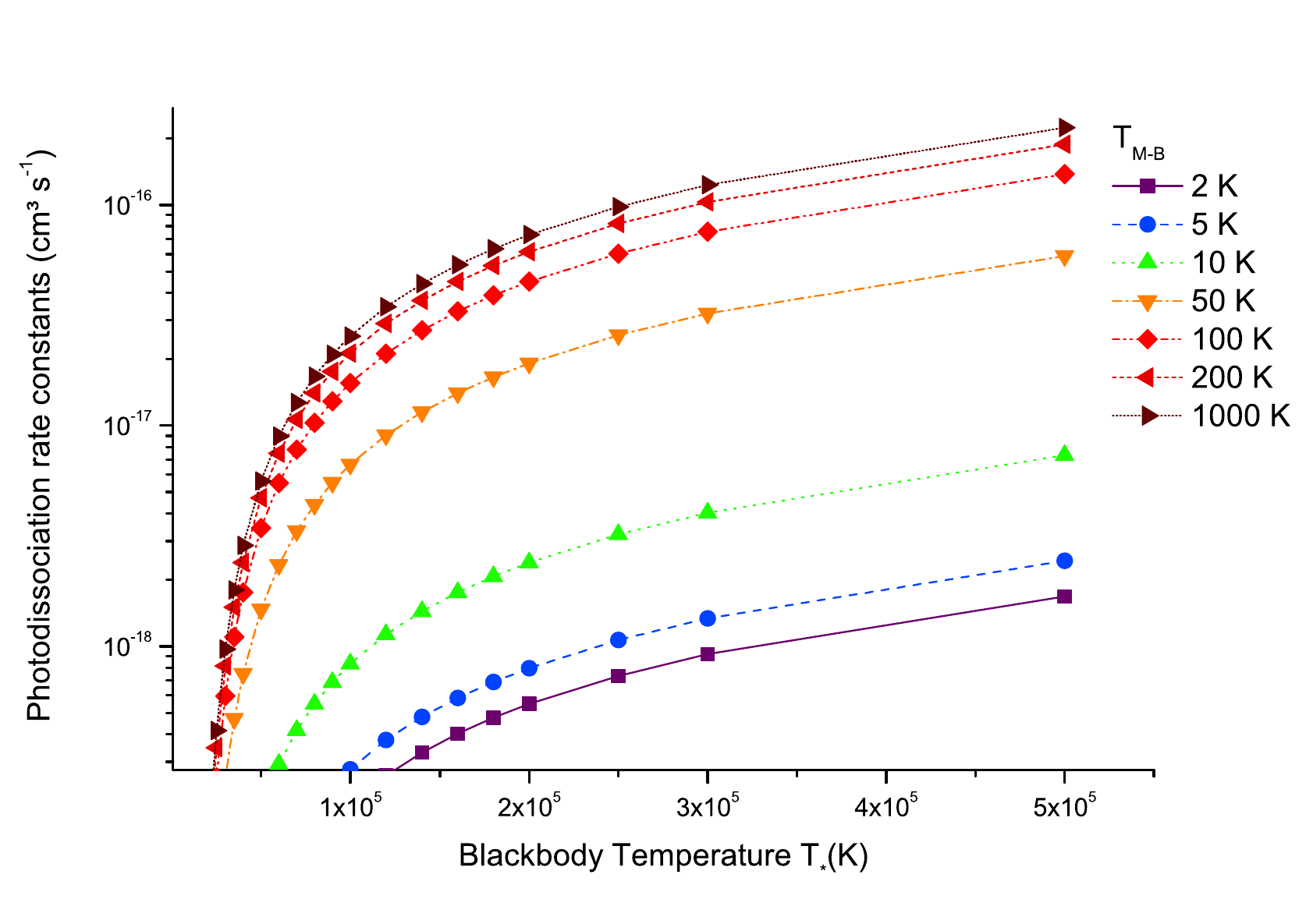}
\caption{Rate constants for photodissociation of the $b$ $^3\mathrm{\Sigma}^+$ state as a function of the temperature $T_*$ of the blackbody, for different Maxwell-Boltzmann distribution temperatures.}
\label{rate_photodissociation}      
\end{figure*}
 As in all calculations of this kind, it should be noted that the photodissociation rate constants are likely to be underestimated at high blackbody temperature due to the absence of excited states beyond $n$ = 3 in our computational basis. 

Likewise, the radiative association rate constants were computed assuming a Maxwell-Boltzmann distribution of the relative kinetic energy of the colliding fragments: 
 
\begin{equation}
k^{\mathrm{RS}} = \sqrt{\frac{8}{\pi\mu k^3_B T}} \int \mathrm{\sigma}^{\mathrm{Rad. Stab.}}_{m J' \rightarrow 0 v'' J''} E_{k} e^{-E_k/k_B T} dE 
\end{equation}
 
Both photodissociation and radiative association rate constants vary strongly with the initial/final vibrational level considered. Qualitatively similar results are to be expected for all systems in which the photodissociation cross sections of the first vibrational levels differ significantly, which depends both on the shapes and on the relative positions of the potential energy curves. Whether neglecting the vibrational dependence of the cross sections is a valid approximation or not is therefore expected to vary from one system to another.

\section{Conclusions}
\label{conclusions}

We have illustrated the main difficulties that can arise when calculating photodissociation cross sections using time dependent methods, particularly when one wants to deduce radiative association cross sections from the same calculations. We used the photodissociation of the $b$ $^3\mathrm{\Sigma}^+$ of HeH$^+$ as an example as this state exhibits several peculiarities, such as a large number of vibrational levels and a relative proximity to several excited electronic states strongly coupled through non-adiabatic interactions. The presence of bound states among those was shown to further complicate matters. Firstly, it leads to the occurrence of Feshbach resonances, which we chose to neglect here since their exact determination through time-dependent approaches is impractical and since the corresponding non-adiabatic transitions are unlikely to occur during radiative association. Secondly, it causes Gibbs oscillations to appear near the threshold energy in partial photodissociation cross sections, which significantly affect the radiative association cross sections at low collision energy.

By combining two methods of computation of the photodissociation cross section, the partial photodissociation cross sections from the thirty-three vibrational levels of the $b$ $^3\mathrm{\Sigma}^+$ state of HeH$^+$ towards the 9 other $^3\mathrm{\Sigma}^+$ and the 6 $^3 \mathrm{\Pi}$ $n$ = 2 and 3 higher-lying electronic states have been calculated with minimal contributions from the Gibbs phenomenon.
Thanks to judicious choices of the propagation time, photodissociation cross sections free from Feshbach resonances were obtained in order to compute the corresponding radiative association cross sections. We have shown that this approach may be more time-efficient than the collisional approach to describe the radiative association towards a specific state for several collisional channels at once. The photodissociation and radiative association rate constants were computed and were shown to be strongly affected by the vibrational dependence of the cross sections. Although it is expected to be smaller, the role of the rotational dependence of the cross sections could be significant due to the contributions of shape resonances.

\ack 

This work was supported by the Communaut\'{e} fran\c{c}aise of Belgium (Action de Recherche Concert\'{e}e) and the Belgian National Fund for
Scientific Research (FRFC and IISN Conventions). S. Vranckx and J. Loreau thank the FNRS for financial support.

\section*{References}
\bibliographystyle{jphysicsB}

\end{document}